\documentclass[a4paper,11pt]{article}
\usepackage{amssymb,amsmath,amsfonts,latexsym,amsthm}
\usepackage[english]{babel}
\usepackage[mathscr]{eucal}
\usepackage{verbatim}
\usepackage{cite}

\date{}
\newtheorem {theorem}{\bf Theorem}
\newtheorem {corol}{\bf Corollary}
\newtheorem {lemma}{\bf Lemma}

\newcommand{\btheorem}{\begin{theorem}}
\newcommand{\etheorem}{\end{theorem}}

\newcommand{\blemma}{\begin{lemma}}
\newcommand{\elemma}{\end{lemma}}

\newcommand{\bcorol}{\begin{corol}}
\newcommand{\ecorol}{\end{corol}}

\newcommand{\bdefin}{\begin{defin}}
\newcommand{\edefin}{\end{defin}}

\newcommand {\bproof }{{\par\medskip\noindent \bf Proof. }}
\newcommand {\eproof }{\hfill $\blacktriangle$ \\ \medskip}

\def\p0{\parindent 0pt}
\begin{document}
\renewcommand{\abstractname}{Abstract}
\begin{center}
\textbf{\Large On weak isometries of Preparata codes}
\vspace*{10pt}

\medskip

\large Ivan Yu. Mogilnykh\\
 \normalsize
\medskip
Sobolev Institute of Mathematics,
 \normalsize
\medskip
 Novosibirsk, Russia

{\it e-mail: \, ivmog84@gmail.com}


\begin{abstract} Let $C_{1}$ and $C_{2}$ be codes with code
distance $d$. Codes $C_{1}$ and $C_{2}$ are called {\it weakly
isometric}, if there exists a mapping $J:C_{1}\rightarrow C_{2}$,
such that for any $x,y$ from $C_{1}$ the equality $d(x,y)=d$ holds
if and only if $d(J(x),J(y))=d$. Obviously two codes are weakly
isometric if and only if the minimal distance graphs of these
codes are isomorphic. In this paper we prove that Preparata codes
of length $n\geq2^{12}$ are weakly isometric if and only if these
codes are equivalent. The analogous result is obtained for
punctured Preparata codes of length not less than $2^{10}-1$.
\end{abstract}

\it{Submitted to Problems of Information Transmission on 11th of
January
 2009.}
\end{center}
\section{Introduction}

Let $E^{n}$ denote all binary vectors of length n. The {\it
Hamming distance} between two vectors from $E^{n}$ is the number
of places where they differ.
 The {\it weight} of vector $x\in E^{n}$ is the distance between this
 vector and the all-zero vector $0^{n}$, and
the {\it support} of $x$ is the set
$supp(x)=\{i\in\{1,\ldots,n\}:x_{i}=1\}$.

 A set $C,$ $C\subset E^{n}$, is called a {\it code} with parameters $(n,M,d)$,
 if $|C|=M$ and the minimal distance between two codewords from $C$ equals $d$.
 We say that a code $C$ is {\it reduced} if it contains all-zero vector.

 A collection of $k$-subsets
(referred to as blocks) of a $n$-set such that any $t$-subset
occurs in $\lambda$ blocks precisely is called a $(\lambda, n, k,
t)$-{\it design}.

The {\it minimal distance graph} of a code $C$ is defined as the
graph with all codewords of $C$ as vertices, with two vertices
 being connected if and only if the Hamming distance between
 corresponding codewords equals to the code distance of the code $C$.
Two codeword of $C$ are called $d$-{\it adjacent} if the Hamming
distance equals code distance $d$ of the code $C$.

Two codes $C_{1}$ and $C_{2}$ of length $n$ are called
 equivalent, if an automorphism $F$ of $E^{n}$ exists such that $F(C_{1})=C_{2}$.
  A mapping $I:C_{1}\rightarrow C_{2}$ of two codes
  $C_{1}$ and $C_{2}$ is called an {\it isometry} between codes $C_{1}$ and $C_{2}$, if the equality $d(x,y)=d(I(x),I(y))$ holds for all $x$ and $y$ from $C_{1}$.
  Then codes $C_{1}$ and $C_{2}$
 are called {\it isometric}.
  A mapping $J:C_{1}\rightarrow C_{2}$
  is called a {\it weak isometry} of codes $C_{1}$ and $C_{2}$ (and codes $C_{1}$ and $C_{2}$
weakly isometrical), if for any $x,y$ from $C_{1}$ the equality
 $d(x,y)=d$ holds if and only if $d(J(x),J(y))=d$ where $d$
  is the code distance of code $C_{1}$. Obviously two codes are weakly isometric if and only if the minimal distance graphs of these codes are isomorphic.
  In \cite{AVG98} Avgustinovich established that any two weakly isometric 1-perfect codes are equivalent. In \cite{OST08}
it was proved that this result also holds for extended 1-perfect
codes.

 In this paper any weak isometry of two Preparata codes (punctured Preparata codes) is proved to be an isometry of these codes.
Moreover, weakly isometric Preparata codes (punctured Preparata
codes) of length $n\geq 2^{12}$ (of length $n\geq 2^{10}-1$
respectively) are proved to be equivalent.

 This topic is closely
related with problem of metrical rigidity of codes. A code $C$ is
called {\it metrically rigid} if any isometry $I:C\rightarrow
E^{n}$ can be extended to an isometry (automorphism) of the whole
space $E^{n}$. Obviously any two metrically rigid isometric codes
are equivalent.
 In \cite{SA03} it was established that any reduced
 binary code of length $n$ containing 2-$(n,k,\lambda)$-design
 is metrically rigid for any $n\geq k^{4}$.

 A maximal binary code of length $n=2^{m}$ for even $m$, $m\geq4$ with code distance
 6 is called a {\it Preparata code} $\overline{P^{n}}$. {\it Punctured Preparata code}
 is a code obtained from Preparata code by deleting one coordinate. By
 $P^{n}$ we denote a punctured Preparata code of length $n$. Preparata codes and
 punctured Preparata codes have some useful properties. All of them are
 distance invariant \cite{SZZ71}, strongly distance invariant
 \cite{VAS}. Also a punctured Preparata code is contained in the unique
 1-perfect code \cite{CZZ73}. An arbitrary punctured Preparata
 code is uniformly packed \cite{SZZ71}. As a consequence of this property,
 codewords of minimal weight of a Preparata code (punctured Preparata code) form a
 design. The last property is crucial in proving the main result
 of  this paper.

\section{Weak isomery of punctured Preparata codes}
\noindent In this section we prove that any two punctured
Preparata codes of length $n$ with isomorphic minimum distance
graphs are isometric. Moreover, these codes are equivalent for
$n\geq 2^{10}-1$. First we give some preliminary statements.
 \blemma \cite{SZZ71}. Let $P^{n}$ be an arbitrary reduced punctured Preparata code.
 Then codewords of weight 5 of the code $P^{n}$ form 2-(n,5,(n-3)/3) design.
 \elemma
Taking into account a structure of the design from this lemma we
obtain

\bcorol Let $P^{n}$ be an arbitrary reduced punctured Preparata
code and $r,s$ be arbitrary elements of the set $\{1,\ldots,n\}$.
 Then there exists exactly one coordinate $t$ such that all codewords of minimal
 weight of the code $P^{n}$ with ones in coordinates $r$ and $s$ has zero in the coordinate $t$.
 \ecorol

 Let
$C$ be a code with code distance $d$ and $x$ be an arbitrary
codeword of $C$ of weight $i$. Denote by $D_{i,j}(x)$
 the set of all codewords of $C$ of weight $j$ which are
 $d$-adjacent with vector $x$. In case when $C$ is a punctured Preparata code we give some
 properties of the set
 $D_{i,j}(x)$ that make the structure of minimal distance graph of
 this code more clear.
 \blemma Let $x$ be an arbitrary codeword of a punctured Preparata code $P^{n}$.
 Then any vector from $D_{i,i-1}(x)$ ($D_{i,i-3}(x)$ и $D_{i,i-5}(x)$
 respectively) has exactly 3 (4 and 5 resp.) zero coordinates
from $supp(x)$ and exactly 2 (1 and 0 resp.)
 nonzero coordinates from $\{1,\ldots,n\}\backslash supp(x)$.
 \elemma
\bproof Suppose a vector $y\in D_{i,i-k}(x)$ has $m_{k}$ zero
coordinates from $supp(x)$. Then it has exactly $m_{k}-k$ nonzero
coordinates from the set $\{1,\ldots,n\}\backslash supp(x)$. Since
$d(x,y)=5$ we have $m_{k}=(5+k)/2$, which implies the required
property for $k=1,3,5$ .
 \eproof

Let $x$ be a codeword of weight $i$ from a $P^{n}$ ;  $m,l$ be
arbitrary coordinates from $supp(x)$. We denote by $A_{m,l}(x)$
$(B_{m,l}(x)$ and $C_{m,l}(x)$) the sets $D_{i,i-1}(x)$
 ($D_{i,i-3}(x)$ and $D_{i,i-5}(x)$ respectively) with coordinates $m$ and $l$ equal to zero.

\blemma Let $x\in P^{n}$, $m,l\in supp(x)$ and $u, v$ be arbitrary
codewords of $P^{n}$ with zeros in coordinates $m$ and $l$ that
are at distance $5$ from $x$. Then $u, v$ do not share zero
coordinates in $supp(x)\setminus \{m,l\}$ and do not share
coordinates equal to one in the set $\{1,\ldots,n\}\backslash
supp(x)$. \elemma
 \bproof Let us suppose the opposite. Then the vectors $x+u$ and $x+v$ of weight five share at least three coordinates with ones in
 them and therefore

  $$d(u,v)=d(x+u,x+v)\leq
 4$$ holds. Since code distance of the code $P^{n}$ equals 5 we get a contradiction. \eproof

 \blemma Let $x$ be an arbitrary codeword of weight $i$ from a punctured Preparata code.
 Then the following inequalities hold:

\begin{equation}\label{lemma4}
(i-3)C_{i}^{2}\leq
3|D_{i,i-1}(x)|+12|D_{i,i-3}(x)|+30|D_{i,i-5}(x)|\leq
(i-2)C_{i}^{2}.
\end{equation}
 \elemma

\bproof Fix two coordinates $m$ and $l$ from $supp(x)$. By Lemma 2
an arbitrary vector from $A_{m,l}(x)$ ($B_{m,l}$ and $C_{m,l}$)
has exactly one zero coordinate (two and three respectively) from
$supp(x)\backslash\{m,l\}$. Then taking into account Lemma 3 the
number of coordinates from $supp(x)\backslash\{m,l\}$ which are
zero for vectors from $A_{m,l}$, $B_{m,l}$
 and $C_{m,l}$ equals $|A_{m,l}(x)|$, $2|B_{m,l}|$ and
 $3|C_{m,l}|$ respectively. Therefore the number of coordinates
 from the $supp(x)\backslash\{m,l\}$ which are zero for  vectors
 from $A_{m,l}\cup B_{m,l}\cup C_{m,l}$ equals

$$|A_{m,l}(x)|+2|B_{m,l}(x)|+3|C_{m,l}(x)|.$$

 Since $x$ is a vector of weight $i$ and $m,l\in supp(x)$, this number does not exceed $i-2$.
 From the other hand by Corollary 1 there exists at most one coordinate from $supp(x)\backslash\{m,l\}$ such that
 all vectors from $A_{m,l}\cup
B_{m,l}\cup C_{m,l}$ have one in it. Thus we have:

$$i-3\leq |A_{m,l}(x)|+2|B_{m,l}(x)|+3|C_{m,l}(x)|\leq i-2.$$
Summing these inequalities for all $m,l\in supp(x)$ we obtain
\begin{equation}\label{once}
(i-3)C_{i}^{2}\leq \sum_{m,l\in supp(x)}|A_{m,l}(x)|+2\sum_{m,l\in
supp(x)}|B_{m,l}(x)|+3\sum_{m,l\in supp(x)}|C_{m,l}(x)|\leq
(i-2)C_{i}^{2}\end{equation}
 As an arbitrary vector from $D_{i,i-1}(x)$ has
exactly 3 zero coordinates from $supp(x)$, any such vector is
counted $C_{3}^{2}$ times in the sum $\sum_{m,l\in
supp(x)}|A_{m,l}(x)|$. Then $$\sum_{m,l\in
supp(x)}|A_{m,l}(x)|=C_{3}^{2}|D_{i,i-1}(x)|.$$ Analogously we
get:
$$\sum_{m,l\in supp(x)}|B_{m,l}(x)|=C_{4}^{2}|D_{i,i-3}(x)|,$$
$$\sum_{m,l\in supp(x)}|C_{m,l}(x)|=C_{5}^{2}|D_{i,i-5}(x)|.$$
So from (\ref{once}) we get (\ref{lemma4}). \eproof

Now we prove the main result using Lemmas 2 and 4.

\btheorem The minimal distance graphs of two punctured Preparata
codes are isomorphic if and only if these codes are isometric.

\etheorem
 \bproof
 It is obvious that if two punctured Preparata codes are isometric
 then they are weakly isometric.

 Let $J:P^{n}_{1}\rightarrow P^{n}_{2}$ be a weak isometry of two
 punctured Preparata codes $P_{1}^{n}$ and $P_{2}^{n}$ of length $n$. Without loss of generality suppose
that $0^{n}\in P^{n}_{1}$, $J(0^{n})=0^{n}$. We now show that
mapping $J$ is an isometry. For proving this it is sufficient to
show that $wt(J(x))=wt(x)$ for all $x\in P^{n}_{1}$.

 Suppose $z$ is a codeword of the code $P^{n}_{1}$, such that $wt(J(z))\neq
 wt(z)=i$ holds and the mapping $J$ preserves weight of all
 codewords of weight smaller that $i$. The vector $z$ satisfying these conditions we call {\it critical}
 Since $J(0^{n})=0^{n}$ and the mapping $J$ preserves the distance
 between all codewords at distance 5, we have $i\geq6$. We prove that there is
 no critical codewords in $P^{n}_{1}$. From $0^{n}\in P^{n}_{1}$ holds that the weak isometry $J$ preserves
 a parity of weight of a vector and therefore $wt(J(z))$ equals either $i+2$ or $i+4$.

   Suppose $wt(J(z))=i+2$. Since $J$ is a weak isometry and $z$ is a critical vector we have the following:
   $|D_{i+2,i-1}(J(z))|=|D_{i,i-1}(z)|$, $|D_{i+2,i-3}(J(z))|=|D_{i,i-3}(z)|$, $|D_{i,i-5}(z)|=0$.
 Taking into account these equalities, from the inequalities of Lemma 4 for vectors $z$ and
 $J(z)$ we get
\begin{equation}\label{lb}
(i-3)C_{i}^{2}\leq 3|D_{i,i-1}(z)|+12|D_{i,i-3}(z)|,
\end{equation}

\begin{equation}\label{ub}
3|D_{i+2,i+1}(J(z))|+12|D_{i,i-1}(z)|+30|D_{i,i-3}(z)|\leq
iC_{i+2}^{2}.
\end{equation}
Multiplying both sides of inequality (\ref{lb}) by $-4$ we get
$$-12|D_{i,i-1}(z)|-48|D_{i,i-3}(z)|\leq-4(i-3)C_{i}^{2}.$$
Summing this inequality with (\ref{ub}) we get
$$3|D_{i+2,i+1(J(z))}|-18|D_{i,i-3}(z)|\leq iC_{i+2}^{2}-4(i-3)C_{i}^{2},$$
 and therefore

\begin{equation}\label{ub2}|D_{i,i-3}(z)|\geq
\frac{4(i-3)C_{i}^{2}-iC_{i+2}^{2}}{18}. \end{equation}

In particular, from the inequality (\ref{ub2}) we have
$|D_{i,i-3}(z)|\geq1$ for $i=6$ and $i=7$. But there is no
codewords of weight 3 and 4 in the $P_{1}$ since $P_{1}$ is
reduced code with code distance 5. Therefore $i\geq8$.
 From Lemma 4 we have the following
 \begin{equation}\label{lb1}|D_{i,i-3}(z)|\leq \frac{(i-2)C_{i}^{2}}{12}.\end{equation}
  But for $i\geq10$ the inequality $3(i-2)C_{i}^{2}<2(4(i-3)C_{i}^{2}-iC_{i+2}^{2})$ holds. This contradicts with (\ref{ub2}) and (\ref{lb1}).

So it is only remains to prove that there are no codewords of
weight 8 and 9, such that their images under the mapping $J$ have
weights 10 and 11 respectively. Obviously the Hamming distance
between any two vectors from $D_{i+2,i-3}(J(z))$  is not less than
6. By Lemma 2 all ones coordinates of each vector from
$D_{i+2,i-3}(J(z))$ are in set $supp(J(z))$. So
$|D_{i+2,i-3}(J(z))|$ does not exceed the cardinality of maximal
constant weight code of length $i+2$, with all code words of
weight being equal $i-3$ and being at distance not less than
 6 pairwise. For $i=8$ and $i=9$ the cardinalities of such codes
 equal to 6 and 11 respectively, but from (\ref{ub2}) we have
 $$|D_{10,5}(J(z))|=|D_{8,5}(z)|\geq12,
|D_{11,6}(J(z))|=|D_{9,6}(z)|\geq21,$$ a contradiction. Therefore
there is no critical vectors $z$ in $P_{1}^{n}$, $wt(z)=i$, such
that $wt(J(z))=i+2$.

   Suppose $wt(J(z))=i+4$. In this case we have
$|D_{i,i-3}(z)|=|D_{i,i-5}(z)|=0$,
$|D_{i+4,i-1}(J(z))|=|D_{i,i-1}(z)|$. Using these equalities we
have from the inequalities of Lemma 4 for the vectors $z$ and
 $J(z)$ the following:
 $$(i-3)C_{i}^{2}\leq3|D_{i,i-1}(z)|,$$

$$ 30|D_{i,i-1}(z)|\leq (i+2)C_{i+4}^{2}.$$
 From these last two inequalities we obtain

$$ \frac{(i-3)C_{i}^{2}}{3}\leq \frac{(i+2)C_{i+4}^{2}}{30},
$$
and therefore

$$ 10i(i-1)(i-3)\leq (i+4)(i+3)(i+2)
$$
 that implies
$$10i(i-1)(i-3)\leq 2i(i+3)(i+2).$$
Since last inequality does not hold for $i\geq6$ there is no critical vectors in $P_{1}^{n}$
 and therefore the mapping $J$ is an isometry.
 \eproof

 In \cite{SA03} the following theorem was proved
 \btheorem Any reduced code of length $n$, that contains a $2-(n,k,\lambda)$-design is metrically rigid for $n\geq k^{4}$. \etheorem

Taking into account that by Lemma 1 any punctured reduced
Preparata code contains
 2-$(n,5,(n-3)/4)$-design applying Theorems 1 and 2 we get

\bcorol Let $n\geq 2^{10}-1$. Two punctured Preparata codes of
length $n$
 are equivalent if and only if the minimal distance graphs of these codes are isomorphic. \ecorol

\section{Weak isometry of Preparata codes}
Using the analogous considerations, Theorems 1,2 and Corollary 2
can easily be extended for extended Preparata codes. We now give
the analogues of Lemmas 1-4 omitting their proofs.

\blemma (\cite{SZZ71}) Let $\overline{P^{n}}$ be an arbitrary
reduced Preparata code.
 Then codewords of weight 6 of code $\overline{P^{n}}$ form 3-(n,6,(n-4)/3)-design.
 \elemma

\blemma Let $x$ be an arbitrary codeword of a Preparata code
$\overline{P^{n}}$, $wt(x)=i$.
 Then any vector from $D_{i,i-2}(x)$ ($D_{i,i-4}(x)$ и $D_{i,i-6}(x)$
 respectively) has exactly 4 (5 and 6 respectively) zero coordinates
from $supp(x)$ and exactly 2 (1 and 0 respectively)
 nonzero coordinates from $\{1,\ldots,n\}\backslash supp(x)$.
 \elemma

\blemma Let $x\in \overline{P^{n}}$, $m,l,k\in supp(x)$, and $u,
v$ be arbitrary codewords of $\overline{P^{n}}$ 
at distance $6$ from the vector $x$ with zero coordinates in
positions $m$, $l$, $k$. Then there is no coordinate from
$supp(x)\setminus \{m,l,k\}$ such that $u, v$ have zeros in it and
there is no coordinate from $\{1,\ldots,n\}\backslash supp(x)$
such that $u,v$ have ones in it. \elemma

 \blemma Let $x$ be an arbitrary codeword of weight $i$ from a Preparata code.
 Then the following inequalities hold:

\begin{equation}
C_{i}^{3}(i-4)\leq
4|D_{i,i-2}(x)|+20|D_{i,i-4}(x)|+60|D_{i,i-6}(x)|\leq
C_{i}^{3}(i-3).
\end{equation}
 \elemma

 Using Lemmas 5-8 and the same arguments as in the proof of Theorem 1
 the following theorem it is not difficult to prove
 \btheorem The minimal distance graphs of two Preparata codes are isomorphic if and only if the codes are isometric. \etheorem

From this theorem, Lemma 5 and Theorem 2 we get
 \bcorol
Let $n\geq 2^{12}$. Two Preparata codes of length $n$
 are equivalent if and only if the minimal distance graphs of these codes are isomorphic.
 \ecorol

The Author is deepfuly grateful to Faina Ivanovna Soloveva for
introducing into the topic, problem statement and all around
support of this work.
 
\end{document}